\documentclass[onecolumn,showpacs,prb]{revtex4}
\usepackage{epsfig}
\usepackage{bm}
\usepackage{amsfonts}
\usepackage{amssymb,amsmath}
\usepackage{enumerate}
\usepackage{multirow}
\usepackage{epsfig}
\usepackage{graphicx}
\usepackage{bm}
\usepackage{hyperref}
\def\l{\left}
\def\r{\right}
\def\be{\begin{equation}}
\def\ee{\end{equation}}
\def\bea{\begin{eqnarray}}
\def\eea{\end{eqnarray}}

\begin{document}

\title{The role of potential in the ghost-condensate dark energy model}

\author{Gour Bhattacharya}
\email{drgour@yahoo.com}
\affiliation{Department of Physics, Presidency University, 86/1, College Street, Kolkata 700 073, India}
%\altaffiliation{Visiting Associate, Inter University Centre for Astronomy and Astrophysics, Pune, India}
\author{Pradip Mukherjee}
\email{mukhpradip@gmail.com}
\affiliation{Department of Physics, Barasat Govt. College, 10 K. N. C. Road, Barasat, Kolkata 700 124, India}
%\altaffiliation{Visiting Associate, Inter University Centre for Astronomy and Astrophysics, Pune, India}
\author{Amit Singha Roy}
\email{singharoyamit@gmail.com}
\affiliation{Department of Physics, Barasat Govt. College, 10 K. N. C. Road, Barasat, Kolkata 700 124, India}
\author{Anirban Saha}
\email{anirban@iucaa.ernet.in}
\affiliation{Department of Physics, West Bengal State University, Barasat, North 24 Paraganas, West Bengal, India}
%\altaffiliation{Visiting Associate, Inter University Centre for Astronomy and Astrophysics, Pune, India}

\begin{abstract}
\noindent 
We consider the ghost-condensate model of dark energy with a generic potential term. The inclusion of the potential is shown to give greater freedom in realising the phantom regime. The self-consistency of the analysis is demonstrated using WMAP7+BAO+Hubble data.
\end{abstract}


\maketitle

\section{Introduction}
Recent cosmological observations % SNIa {\cite{c1}}, WMAP {\cite{c2}}, SDSS {\cite{c3}} and X-ray {\cite{c4}} 
indicate late-time acceleration of the observable universe \cite{NL1, NL2}. Why the evolution of the universe is interposed between an early inflationary phase and the late-time acceleration is a yet-unresolved problem. Various theoretical attempts have been undertaken to confront this observational fact. Although the simplest way to explain this behavior is the consideration of a cosmological constant \cite{wein}, the known fine-tuning problem \cite{DE} led to the dark energy paradigm. Here one introduces exotic dark energy component in the form of scalar fields such as quintessence \cite{quint1, quint2, quint3, quint4, quint5, quint6, quint7}, k-essence \cite{kessence1, kessence2, kessence3} etc. Quintessence is based on scalar field models using a canonical field with a slowly varying potential. On the other hand the models grouped under k-essence are characterized by noncanonical kinetic terms. 
%given by $P\l(X, \phi\r)$ in the scalar field action where $\phi$ is the scalar field and $X = -\frac{1}{2}g^{\mu \nu}\partial_{\mu}\phi \partial_{\nu}\phi$ is its kinetic energy. 
%The central point of the models grouped under 
A key feature of the k-essence models is that the cosmic acceleration is realized by the kinetic energy  of the scalar field. The popular models under this category include the phantom model, the ghost-condensate model etc \cite{DE, bamba}. 

     It is well-known that the late time cosmic acceleration requires an exotic equation of state $\omega_{DE} < -\frac{1}{3}$. From the seven year Wilkinson Microwave Anisotropy Probe (WMAP7) observations data, distance measurements from the BAO and the Hubble constant measurements the value of a constant EOS for dark energy has been estimated as $\omega_{DE} = -1.10 \pm 0.14 \left(68\% {\rm CL}\right)$ for flat universe \cite{Komatsu}. Primary results from PAN-STARRS in fact pushes this limit further \cite{PANSTARR} though the full data is yet to arrive. No scalar field dark energy model with canonical kinetic energy term can achieve $\omega_{DE} < -1$. For this one has to consider a scalar field theory with negative kinetic energy along with a field potential. The resulting phantom model \cite{phantom1, phantom2, phantom3, phantom4, phantom5, phantom6} is extensively used to confront cosmological observation \cite{phantom_obs1, phantom_obs2, phantom_obs3, phantom_obs4, phantom_obs5, phantom_obs6}. 
%Recently this WMAP7+BAO+Hubble data has been confronted successfully with the phantom model \cite{gumjudpai}.
     
                   The phantom model is however ridden with various instabilities as its energy density is unbounded. This instability can be eliminated in the so-called ghost-condensate (GC) models \cite{GC} by including a term quadratic in the kinetic energy. In this context let us note that to realize the late-time acceleration scenario some self-interaction must be present in the phantom model. In contrast, in the GC models the inclusion of self-interaction potential of the scalar field is believed to be a matter of choice \cite{DE}. This fact, though not unfamiliar, has not been emphasised much in the literature. In the present paper we show that by including a potential term in the GC model brings more flexibility in realising the phantom evolution. 

	It is well-known that the GC model without the potential resides within the phantom regime for a certain range of values of the scalar field kinetic energy \cite{DE}. We will demonstrate here that these range is widened in presence of a generic potential term. %Consequently some new conditions will emarge. 
Note that this widening is a consequence of the field theoretic aspects of the present dark energy model. Also it crucially depends on the positive energy condition. The question arises whether these conditions %(i.e., the positive energy condition and the allowed range of the kinetic energy) identified in this paper 
for achieving the phantom regime are consistent with the scalar field dynamics or not. 

Now the scalar field dynamics is not independent but is coupled with gravity. Usually one assumes a specific potential and the consequent evolution is studied. But in this paper our objective is to point out the advantage of including a potential in the GC model for achieving the phantom regime. Thus we start with an arbitrary potential and exploit a specific feature of the GC action to show that the potential can be expressed in terms of observable parameters (e.g. $\dot{H}$) once the evolution of the scale factor is chosen. Naturally we use the phantom power law here. Consequently the kinetic and potential energy are expressed as functions of time. We still require observational data to fix the geometric parameters appearing in these functional relations so that their time-evolutions can be explicitly obtained.
%Of course this question will only be meaningful when actual observational data are used to fix the the necessary geometric parameters. 
For this purpose the combined WMAP7+BAO+Hubble data will be used. The potential and kinetic energy are plotted. The plots clearly show that the criteria derived here for our model to realize the phantom evolution hold throughout the entire late-time evolution.

The organization of this paper is as follows. In section II we briefly review the ghost-condensate model with an arbitrary potential. The equations of motion for the scalar field and the scale factor are derived. These equations exhibit the coupling between the scalar field dynamics and gravity. Expressions for the energy density and pressure of the dark energy components are computed. These expressions are used in section III to find the criteria for the model to acquire phantom evolution. In section IV we utilize an obvious algebric consistency which leads to a quadratic equation in the potential. Solving this the generic potential is expressesed in terms of measurable geometric quantities. To fix these geometric quantities phantom power law evolution is assumed and the combined WMAP7+BAO+Hubble data is used  in section V. The explicite time variations of the potential and the kinetic energy are obtained. We provide the plots of these quantities throughout the late-time evolution. Remarkably the conditions for the phantom regime given in section III are observed to hold. FInally we conclude in section VI.

%%%%%%%%%%%%%%%%%%%%%%%%%%%%%%%%%%%%%%%%%%%%%%%%%%%%%%%%%%%%%%%%%%%%%%%%%%%%%%%%%%%
\section{The model}
\label{model}
%%%%%%%%%%%%%%%%%%%%%%%%%%%%%%%%%%%%%%%%%%%%%%%%%%%%%%%%%%%%%%%%%%%%%%%%%%%%%%%%%%%
In this section we consider the ghost-condensate model with a self-interaction potential $V(\phi)$.
The action is given by 
\begin{equation}
S=\int d^{4}x \sqrt{-g} \left[\frac{R}{2k^{2}}
+{\cal{L}}_{\phi}
+{\cal{L}}_{\rm{m}}\right], \label{ghost}
\end{equation}
where 
\begin{eqnarray}
{\cal{L}}_{\phi} &=& -X + \frac{X^{2}}{M^{4}} - V\left( \phi \right) \label{l}\\
X &=& -\frac{1}{2}g^{\mu \nu}\partial_{\mu}\phi \partial_{\nu}\phi
\label{kinetic}
\end{eqnarray}
%$V(\phi)$ is the ghost-condensate field potential, 
$M$ is a mass parameter, $R$ the Ricci scalar and $G = k^{2}/8\pi$ the gravitational constant. The term ${\cal{L}}_{\rm{m}}$ accounts for the total (dark plus baryonic) matter content of the
universe, which is assumed to be a barotropic fluid with energy density $\rho_m$ and pressure $p_m$, and equation-of-state parameter $w_m=p_m/\rho_m$. We neglect the radiation sector for simplicity.% since only small redshifts are  considered.

%In this section we present ghost condensate cosmology under power-law expansion.
 The action given by equation (\ref{ghost}) describes a scalar field interacting with gravity. 
%We concentrate first on the Einstein's equations following from (\ref{ghost}). 
Invoking the cosmological principle one requires the metric to be 
%homogenous and isotropic 
of the Robertson-Walker (RW) form
\begin{equation}
ds^2=dt^2-a^2(t)\left[\frac{dr^2}{1-Kr^2}+r^2d\Omega_2^2\right],
\end{equation}
where $t$ is the cosmic time, $r$ is the spatial radial coordinate, $\Omega_2$ is the 2-dimensional unit sphere volume, $K$ characterizes the curvature of 3-dimensional space and
%, $K=-1,0,1$ corresponds to open, flat and closed universe respectively. 
$a(t)$ is the scale factor.
The Einstein equations lead to the Freidman equations 
\begin{eqnarray}
H^{2}&=&\frac{k^{2}}{3}\Big(\rho_{m}+\rho_{\phi}\Big)-
\frac{K}{a^2} \label{FR1}\\
\dot{H}&=&-\frac{k^{2}}{2} \Big(\rho_{m}+p_m+\rho_{\phi}+p_{\phi}\Big)+\frac{K}{a^2}, \label{FR2} 
\end{eqnarray}
In the above a dot denotes derivative with respect to $t$ and $H\equiv\dot{a}/a$ is the Hubble parameter. In these expressions, $\rho_{\phi}$ and $p_\phi$  are respectively the energy density
and pressure of the scalar field. The quantities $\rho_{\phi}$ and $p_\phi$ are defined through the symmetric energy-momentum tensor
\begin{equation}
T^{(\phi)}_{\mu\nu} = \frac{-2}{\sqrt{-g}}\frac{\delta}{\delta g^{\mu\nu}}\left({\sqrt{-g}} \right)
\end{equation} 
A straightforward calculation gives
\begin{equation}
T^{(\phi)}_{\mu\nu} = g_{\mu\nu}{\cal{L}}_{\phi} + \left(-1 + \frac{2X}{M^4}\right)\partial _\mu\phi\partial _\nu\phi
\end{equation}
Assuming a perfect fluid model we identify
\begin{eqnarray}
\rho_{\phi} &=& -X + \frac{3X^{2}}{M^{4}} + V\left( \phi \right) 
\label{density}\\
p_{\phi} &=& {\cal{L}}_{\phi} = -X + \frac{X^{2}}{M^{4}} - V\left( \phi \right) 
\label{pressure}
\end{eqnarray}
The equation of motion for the scalar field $\phi$ can be derived from the action (\ref{ghost}). Due to the isotropy of the FLRW universe the scalar field is a function of time only. Consequently, its equation of motion reduces to
\begin{equation}
\l(1 - \frac{3 \dot{\phi}^{2}}{M^{4}}\r)\ddot{\phi} + 3H\l(1 - \frac{\dot{\phi}^{2}}{M^{4}}\r)\dot{\phi}- \frac{dV}{d \phi} = 0. \label{eqm}
\end{equation}
As is well known the same equation of motion follows from the conservation of $T_{\mu\nu}$. Indeed under isotropy the equations (\ref{density}) and (\ref{pressure}) reduce to 
\begin{eqnarray}
\rho_{\phi} &=& - \frac{1}{2}\dot{\phi}^{2} + \frac{\dot{3\phi}^{4}}{4 M ^{4}} + V\left( \phi \right)
\label{rhophi}\\
p_{\phi} &=& - \frac{1}{2}\dot{\phi}^{2} + \frac{\dot{\phi}^{4}}{4 M ^{4}} - V\left( \phi \right)
\label{pphi}
\end{eqnarray}
From the conservation condition $\nabla_\mu T^{(\phi)\mu\nu} = 0$ we get 
\begin{equation}
\dot{\rho}_\phi+3H(\rho_\phi+p_\phi)=0, \label{rhodot}
\end{equation}
which, written equivalently in field terms gives equation (\ref{eqm}).

To complete the set of differential equations (\ref{FR1}), (\ref{FR2}), (\ref{rhodot}) we include the equation for the evolution of matter density
\begin{eqnarray}
\dot{\rho}_m+3H(1+w_m)\rho_m=0, \label{rhomdot}
\end{eqnarray}
where $w_{m}=p_{m}/\rho_{m}$ is the matter equation of state parameter. The solution to equation (\ref{rhomdot}) can immediately be written down as
\begin{equation}
\frac{\rho_m}{\rho_{m0}} = \l[\frac{a\l(t_{0}\r)}{a\l(t\r)}\r]^{n}, \label{rhom}
\end{equation}
where $n = 3 (1 + w_m) $ and $\rho_{m0} \geq 0$ is the value of matter density at present time $t_0$. Now, the set of equations (\ref{FR1}), (\ref{FR2}), (\ref{rhodot}) and (\ref{rhomdot}) must give the dynamics of the scalar field under gravity in a self-consistent manner. In the next section we investigate the criteria for the GC model to realise the phantom evolution.

%%%%%%%%%%%%%%%%%%%%%%%%%%%%%%%%%%%%%%%%%
\section{Criteria for Realising the phantom regime}
%%%%%%%%%%%%%%%%%%%%%%%%%%%%%%%%%%%%%%%%% 
The phantom regime is demarcated by $\omega_{\phi} < -1$ where $\omega_{\phi}$ is the dark energy equation of state (EoS) parameter defined as
\begin{eqnarray}
{\omega_\phi}&=&\frac{P_\phi}{\rho_\phi}
\label{omega}
\end{eqnarray}
In the present section we investigate the criteria for our model to be in the phantom regime using the definition (\ref{omega}) only without recourse to actual dynamics. From equation (\ref{rhophi}) and (\ref{pphi}), the EoS parameter for the  field $\phi$ is obtained as 
\begin{eqnarray}
{\omega_\phi}&=&\frac{-\frac{\dot{\phi}^{2}}{2}+ \frac{\dot{\phi}^{4}}{4 M^{4}}-V(\phi)}{-\frac{\dot{\phi}^{2}}{2} + \frac{3\dot{\phi}^{4}}{4 M^{4}} + V(\phi)}
%\nonumber\\ &=&-\frac{V{(\phi)}+f{(\dot{\phi})}+\frac{\dot{\phi}^4}{2M^4}}{V{(\phi)}-f{(\dot{\phi})}}
\label{omegaphi}
\end{eqnarray}
Defining $f\left(\dot{\phi}\right) = \left(\frac{\dot{\phi}^{2}}{2} - \frac{3\dot{\phi}^{4}}{4 M^{4}}\right)$ (\ref{omegaphi}) can be cast in the form 
\begin{eqnarray}
{\omega_\phi} =-1 -\frac{\dot{\phi}^{2}\left(1 - \frac{\dot{\phi}^{2}}{M^{4}}\right)}{V\left(\phi\right) - f(\dot{\phi})}
\label{omegaphi1}
\end{eqnarray}
This equation is more suitable to discuss the conditions for achieving the phantom regime. 
\begin{enumerate}
\item
First assume that there is no self-interaction, i.e., $V\left(\phi\right) = 0$. The positive energy condition ensures that $\rho_{\phi} = -f(\dot{\phi}) > 0$. Thus, for $\omega_{\phi} < - 1$ we require $\left(1 - \frac{\dot{\phi}^{2}}{M^{4}}\right)>0$. These lead to the following bounds \cite{DE} 
\begin{eqnarray}
\frac{2}{3}M^4 < \dot{\phi}^{2} < M^{4}
\label{criteria_1}
\end{eqnarray} 
so that the phantom regime is attained.
\item
Now suppose, $V\left(\phi\right) \ne 0$. From the positive energy condition $\rho_{\phi} > 0$, (see equation (\ref{rhophi})) we get  
\begin{center}
\begin{eqnarray}
V\left(\phi\right) > f(\dot{\phi}) 
\label{cond_1}
\end{eqnarray} 
\end{center} 
The only restriction imposed is now $\dot{\phi}^{2} < M^{4}$. Of course $\dot{\phi}$ is real so we now require 
\begin{center}
\begin{eqnarray}
0 < \dot{\phi}^{2}< {M^4}
\label{cond_2}
\end{eqnarray}
\end{center} 
\end{enumerate}
Comparing the equations (\ref{criteria_1}) and (\ref{cond_2}) it is clear that inclusion of appropriate self-interaction provides greater flexibility to realise the phantom domain. In the phantom domain the scale factor evolves according to the phantom power law \cite{DE}:
\begin{eqnarray}
a(t)={a_0}{\left(\frac{t_s-t}{t_s-t_0}\right)}^{\beta}
\label{phantom_power_law}
\end{eqnarray}
where $t_{0}$ and $t_{s}$ are the present time and big-rip time \cite{phantom1, phantom2} respectively. These parameters are obtained from observational data. In this connection it is important to note that the condition (\ref{cond_2}) is obtained from the definition of the EoS (\ref{omega}) which in turn follows from the particular energy-momentum tensor obtained from the model (\ref{l}, \ref{kinetic}). Such quantities have been termed as the `physical variables' in the literature \cite{sahni}. In contrast the geometric quantities (e.g. the Hubble parameter $H\left(t \right)$ and its time-derivative) are determined from observations in a model-independent way \cite{sahni}. Naturally one wonders whether the dynamical evolution of the system according to the phantom power law always conforms with the condition (\ref{cond_2}). 

At this point, one should note that in general, the dynamical evolution of the fields can not be worked out if the potential is not specified. However, as emphasised in the introduction, a specific aspect fo the GC model (\ref{l}, \ref{kinetic}) allows us to express the arbitrary potential in terms of geometric quantities. Consequently, the field variables and the potential here can be expressed as function of time once the geometric parameters involved in (\ref{phantom_power_law}) are fixed from observational data. It will then be possible to answer whether our criteria remains satisfied with the phantom evolution throughout the late time. 

%%%%%%%%%%%%%%%%%%%%%%%%
\section{The potential from geometric quantities}
%%%%%%%%%%%%%%%%%%%%%%%%
In this section we will exploit the structure of the model (\ref{l}, \ref{kinetic}) to establish an algebraic identity which will enable us to express the generic potential in terms of geometric quantities. We start by constructing two independent combinations of the pressure and energy density of the dark energy sector in terms of the Hubble parameter $H$, matter energy density $\rho_{m}$, matter equation of state parameter $w_{m}$ and curvature parameter $K$ using (\ref{FR1}), (\ref{FR2}) and (\ref{rhomdot})
\begin{eqnarray}
\rho_{\phi} + p_{\phi} =A &=& -\frac{2 \dot{H}}{k^{2}}-\frac{n}{3} \rho_{m}  + \frac{2K}{k^{2}a^{2}} \label{A} \\
\rho_{\phi} + 3p_{\phi} =B &=& -\frac{6 \ddot{a}}{k^{2}a} - \l(n-2\r) \rho_{m} \label{B} 
\end{eqnarray}
Using equations (\ref{rhophi}) and (\ref{pphi}), we rewrite these combinations in terms of the ghost-condensate field derivative $\dot{\phi}$ and potential $V\left( \phi \right)$:
\begin{eqnarray}
\rho_{\phi} + p_{\phi} =A &=&  - \dot{\phi}^{2} + \frac{\dot{\phi}^{4}}{M ^{4}} \label{A1} \\
\rho_{\phi} + 3p_{\phi} =B &=&  - 2 \dot{\phi}^{2} + \frac{3\dot{\phi}^{4}}{2M ^{4}} - 2 V\left( \phi \right) \label{B1} 
\end{eqnarray}
Inverting the equations (\ref{A1}, \ref{B1}) we can write $\dot{\phi}^{2}$ and $\dot{\phi}^{4}$ in terms of $A$, $B$ and $V\left( \phi \right)$ as 
\begin{eqnarray}
\dot{\phi}^{2} = 3A - 2B - 4V\l(\phi\r)
\label{A2} \\
\dot{\phi}^{4} = 2M^{4}\l[\l(2A - B\r) - 2V\l(\phi\r)\r]
\label{B2}
\end{eqnarray}
Note that there is an obvious suggestion lurking behind the equations (\ref{A2}) and (\ref{B2}), namely, the algebraic identity
\begin{eqnarray}
\l(\dot{\phi}^{2}\r)^{2} = \dot{\phi}^{4}
\label{identity}
\end{eqnarray}
If one substitutes both sides of the identity from equations (\ref{A2}) and (\ref{B2}) an equation is obtained which contains only geometric quantities, except for the potential. Thus it allows us to express the arbitrary potential in terms of these geometric quantities. Note further that the statement holds because we have already agreed to assume the phantom power law with the geometric parameters appearing in it fixed by observational data. This is clearly the unique feature of the ghost-condensate model (\ref{l}, \ref{kinetic}) which has been referred to in the above. 

Utilizing the identity we obtain the following equation quadratic in $V\l(\phi\r)$
\begin{eqnarray}
V^{2}\l(\phi\r) && + \l(B - \frac{3A}{2} + \frac{M^{4}}{4}\r) V\l(\phi\r) \nonumber \\
&& + \frac{\l(3A - 2B\r)^{2} - 4 M^{4}\l(A - B/2\r)}{16} = 0
\label{quad} 
\end{eqnarray}

At this point one may ask whethar the constraining equation (\ref{quad}) on $ V\l(\phi\r) $ at all allows a real solution. Solving (\ref{quad}) we get 
\begin{eqnarray}
V\l(\phi\r) = \l(\frac{3A - 2B}{4} -\frac{M^{4}}{8}\r) \pm \l\{\frac{M^{4}}{16}\l(\frac{M^{4}}{4} +A\r)\r\}^{\frac{1}{2}} 
\label{V} 
\end{eqnarray}
The reality condition is thus 
\begin{eqnarray}
\l(\frac{M^{4}}{4} +A \r) \ge 0     
\label{reality} 
\end{eqnarray}
That this condition is satisfied in general can be established explicitly if we substitute for $A$ from equation (\ref{A1}) which gives 
\begin{eqnarray}
\l(\frac{M^{4}}{4} +A \r)  = \frac{1}{M^{4}}\l(\dot{\phi}^{2} - \frac{M^{4}}{2}\r)^{2} \ge 0
\label{reality_check} 
\end{eqnarray}
Since from physical consideration the interaction potential is required to be real the above observation indicates the consistency of our formalism. 

In the next section we will utilize the solution (\ref{V}) to express the geneic potential as a function of time employing the phantom power law. This is the point of departure of our work from the existing works with the GC model available in the literature. This, as has been explained in the above, suits our purpose of showing that inclusion of a potential widens the allowed range of kinetic energy of the GC model to realise the phantom regime. Needless to say it is imparetive to varify that the criteria identified above are consistent with the dynamical evolution.

%%%%%%%%%%%%%%%%%%%%%%%%%
\section{Our model and the phantom evolution}
%%%%%%%%%%%%%%%%%%%%%%%%%
 In this section we will verify the validity of the criteria (\ref{cond_1}, \ref{cond_2}) in the phantom evolution scenario. Assuming a phantom power law the time evolutions of both the potential and kinetic energies will be studied. To get explicit time variations of these quantities we require the values of various parameters appearing therein. These parameters include the phantom power law exponent, the big rip time as well as the present values of energy density etc. We use the combined WMAP7+BAO+Hubble as well as WMAP7 data \cite{Komatsu} as standard data set \cite{gumjudpai}. Also in our model there is a free parameter $M$, the value of which will be estimated self-consistently using the same ovservational data. 

%%%%%%%%%%%%%%%%%%%%%%%%%%%%%%%%%%%%%%%%%%%%%%%%%
\subsection{Consequence of the phantom power law}
%%%%%%%%%%%%%%%%%%%%%%%%%%%%%%%%%%%%%%%%%%%%%%%%%
We will now find explicit expressions of the potential and kinetic energies as functions of time. The potential is already given by equation (\ref{V}). Substituting $A$ and $B$ from (\ref{A}) and (\ref{B}) respectively we get\footnote{Note that we are assuming dust matter and flat geometry.} 
\begin{eqnarray}
V(\phi)=\left[\frac{3H^2}{8{\pi}G}+\frac{3\dot{H}}{16{\pi}G}-\frac{\rho_m}{4}-\frac{M^4}{8}\right]\nonumber\\
\pm\frac{1}{2}\left[\left(-\frac{\dot{H}}{4{\pi}G}-{\rho_m}+\frac{M^4}{4}\right)\frac{M^4}{4}\right]^\frac{1}{2}
\label{V1}
\end{eqnarray}
Now solving equation (\ref{A1}) kinetic energy term is obtained as 
\begin{eqnarray}
\dot{\phi^2}=\frac{M^4}{2}\mp\left[\left(-\frac{\dot{H}}{4{\pi}G}-{\rho_m}+\frac{M^4}{4}\right){M^4}\right]^\frac{1}{2}\label{KE1}                                                                                                                                                      
\end{eqnarray}
The choice of signs in the equations (\ref{V1}) and (\ref{KE1}) should be noted. This choice is done so as to satisfy (\ref{B1}).

Using the phantom power law we find 
\begin{eqnarray}H &=& -\frac{\beta}{t_s-t} \label{H}\\
\dot{H} &=& -\frac{\beta}{(t_s-t)^2}\label{Hdot}
\end{eqnarray} 
Substituting these and restoring to S.I. units eqation (\ref{V1}) and (\ref{KE1}) become,
\begin{eqnarray}
V(t)=&\left[\frac{3{\beta^2}{c^2}}{8{\pi}G({t_s}-t)^2}-\frac{3{\beta}{c^2}}{16{\pi}G({t_s}-t)^2}-\frac{1}{4}\frac{{\rho_{m0}}{c^2}}{{a_0}^3}{\left(\frac{{t_s}-{t_0}}{{t_s}-{t}}\right)^{3{\beta}}}-\frac{{M_{S.I.}}^4}{8}\right]\nonumber\\
&\pm\frac{1}{2}\left[\left(\frac{{\beta}{c^2}}{4{\pi}G({t_s}-t)^2}-\frac{{\rho_{m0}}{c^2}}{{a_0}^3}\left(\frac{{t_s}-{t_0}}{{t_s}-{t}}\right)^{3{\beta}}+\frac{{M_{S.I.}}^4}{4}\right)\frac{{M_{S.I.}}^4}{4}\right]^\frac{1}{2}
\label{V_f}\end{eqnarray}
and
\begin{eqnarray}
{\dot\phi^2}=\frac{{M_{S.I.}}^4}{2}\mp\left[\left(\frac{{\beta}{c^2}}{4{\pi}G({t_s}-t)^2}-\frac{{\rho_{m0}}{c^2}}{{a_0}^3}\left(\frac{{t_s}-{t_0}}{{t_s}-{t}}\right)^{3{\beta}}+\frac{{M_{S.I.}}^4}{4}\right){{M_{S.I.}}^4}\right]^\frac{1}{2}
\label{KE_f}
\end{eqnarray}
where ${M_{S.I.}}={M}$(in ev)$\times1.62\times{10^{-2}}.$ 
%${\dot{a}}=-{\frac{\beta}{t_s-t}}a$,
Equations (\ref{V_f}) and (\ref{KE_f}) are the desired time variations of the potential and kinetic energies if the phantom power law is imposed. 

 To proceed further input from the observational data is required. This will enable us to determine  the values of the different geometric parameters appearing in the expressions of above (\ref{V_f}, \ref{KE_f}). It will then be possible to check the validity of the conditions (\ref{cond_1}, \ref{cond_2}). However, before invoking the observational data a consistency check is necessary. This involves the verification whether the reconstructed potential and kinetic energy (\ref{V_f}, \ref{KE_f}) satisfy equation (\ref{eqm}), the eqution of motion of the scalar field. The necessary calculations for the consistency check will be given in the next subsection.
%%%%%%%%%%%%%%%%%%%%%%%%
\subsection{A consistency check}
%%%%%%%%%%%%%%%%%%%%%%%%
Since $V$ is given as function of $t$ we use the chain rule of differentiation to write
\begin{equation}
\frac{dV}{dt} = \frac{1}{\dot{\phi}}\frac{dV}{d\phi}
\end{equation}
Substituting this in (\ref{eqm}) and after a few steps of calculation we get the equivalent of (\ref{eqm}) as
\begin{eqnarray}
%\left(1-\frac{3{{\dot{\phi}}^2}}{{M^4}}\right){\ddot{\phi}}+3H\left(1-\frac{{{\dot{\phi}}^2}}{{M^4}}\right){\dot{\phi}}-\frac{dV}{d \phi} = 0\nonumber\\
%i.e;  \left(1-\frac{3{{\dot{\phi}}^2}}{{M^4}}\right){\ddot{\phi}}+3H\left(1-\frac{{{\dot{\phi}}^2}}{{M^4}}\right){\dot{\phi}}-\frac{1}{\dot{\phi}}\frac{dV}{dt} = 0\nonumber\\
%i.e; \left(1-\frac{3{{\dot{\phi}}^2}}{{M^4}}\right){{\ddot{\phi}}{\dot{\phi}}}+3H\left(1-\frac{{{\dot{\phi}}^2}}{{M^4}}\right){{\dot{\phi}}^2}-\frac{dV}{dt} = 0\nonumber\\
 \frac{1}{2}\frac{d{\dot{\phi}^2}}{dt}-\frac{3}{4{M^4}}\frac{d{\dot{\phi}^4}}{dt}+3H\left(1-\frac{{{\dot{\phi}}^2}}{{M^4}}\right){{\dot{\phi}}^2}=\frac{dV}{dt}\label{eqm1}
\end{eqnarray}
The above form of (\ref{eqm}) can be readily used to verify whether the reconstructed potential (\ref{V_f}) %and kinetic energy (\ref{KE_f}) are 
is consistent with the equation of motion for $\phi$.
Using (\ref{KE_f}) in the left hand side (L.H.S) of (\ref{eqm1}) we get
\begin{eqnarray}
{\rm L.H.S.}
%\frac{1}{2}\frac{d{\dot{\phi}^2}}{dt}-\frac{3}{4{M^4}}\frac{d{\dot{\phi}^4}}{dt}+3H\left(1-\frac{{{\dot{\phi}}^2}}{{M^4}}\right){{\dot{\phi}}^2}\nonumber\\
&=&\mp\frac{1}{4}\frac{\left[\left(\frac{{\beta}{c^2}}{2{\pi}G({t_s}-t)^3}-\frac{3{\rho_{m0}}{\beta{c^2}}}{{{a_0}^3}(t_s-t)}{\left(\frac{{t_s}-{t_0}}{{t_s}-{t}}\right)^{3{\beta}}}\right){{M_{S.I.}}^4}\right]}{\left[(\frac{{\beta}{c^2}}{4{\pi}G({t_s}-t)^2}-\frac{{\rho_{m0}}{{c^2}}}{{{a_0}^3}}{\left(\frac{{t_s}-{t_0}}{{t_s}-{t}}\right)^{3{\beta}}}+\frac{{M_{S.I.}}^4}{4}){{M_{S.I.}}^4}\right]^{\frac{1}{2}}}
\nonumber\\
&-&\frac{3}{4}{\left[\left(\frac{{\beta}{c^2}}{2{\pi}G({t_s}-t)^3}-\frac{3{\rho_{m0}}{\beta{c^2}}}{{{a_0}^3}(t_s-t)}{\left(\frac{{t_s}-{t_0}}{{t_s}-{t}}\right)^{3{\beta}}}\right)\right]}\nonumber\\
&&\pm\frac{3}{8}\frac{\left[\left(\frac{{\beta}{c^2}}{2{\pi}G({t_s}-t)^3}-\frac{3{\rho_{m0}}{\beta{c^2}}}{{{a_0}^3}(t_s-t)}{\left(\frac{{t_s}-{t_0}}{{t_s}-{t}}\right)^{3{\beta}}}\right){{M_{S.I.}}^4}\right]}{\left[(\frac{{\beta}{c^2}}{4{\pi}G({t_s}-t)^2}-\frac{{\rho_{m0}}{{c^2}}}{{{a_0}^3}}{\left(\frac{{t_s}-{t_0}}{{t_s}-{t}}\right)^{3{\beta}}}+\frac{{M_{S.I.}}^4}{4}){{M_{S.I.}}^4}\right]^{\frac{1}{2}}}\nonumber\\
&&+\frac{3{\beta^2}{c^2}}{4{\pi}G({t_s}-t)^3}-\frac{3{\rho_{m0}}{\beta{c^2}}}{{{a_0}^3}(t_s-t)}{\left(\frac{{t_s}-{t_0}}{{t_s}-{t}}\right)^{3{\beta}}}
\end{eqnarray}
Arranging terms, this may be re-expressed as
\begin{eqnarray}
{\rm L.H.S.}&=&\left[\frac{3{\beta^2}{c^2}}{4{\pi}G({t_s}-t)^3}-\frac{3{\beta}{c^2}}{8{\pi}G({t_s}-t)^3}-\frac{3}{4}\frac{{\rho_{m0}}{\beta{c^2}}}{{{a_0}^3}(t_s-t)}{\left(\frac{{t_s}-{t_0}}{{t_s}-{t}}\right)^{3{\beta}}}\right]\nonumber\\
&&\pm\frac{1}{8}\frac{\left[\left(\frac{{\beta}{c^2}}{2{\pi}G({t_s}-t)^3}-\frac{3{\rho_{m0}}{\beta{c^2}}}{{{a_0}^3}(t_s-t)}{\left(\frac{{t_s}-{t_0}}{{t_s}-{t}}\right)^{3{\beta}}}\right){{M_{S.I.}}^4}\right]}{\left[(\frac{{\beta}{c^2}}{4{\pi}G({t_s}-t)^2}-\frac{{\rho_{m0}}{{c^2}}}{{{a_0}^3}}{\left(\frac{{t_s}-{t_0}}{{t_s}-{t}}\right)^{3{\beta}}}+\frac{{M_{S.I.}}^4}{4}){{M_{S.I.}}^4}\right]^{\frac{1}{2}}}\label{rescale}
\end{eqnarray}
Now
%The right hand side of (\ref{eqm1}) is obtained
 from a straightforward differentiation of $V(t)$ using (\ref{V_f})we derive
\begin{eqnarray}
\frac{dV}{dt}
&=&\left[\frac{3{\beta^2}{c^2}}{4{\pi}G({t_s}-t)^3}-\frac{3{\beta}{c^2}}{8{\pi}G({t_s}-t)^3}-\frac{3}{4}\frac{{\rho_{m0}}{\beta{c^2}}}{{{a_0}^3}(t_s-t)}{\left(\frac{{t_s}-{t_0}}{{t_s}-{t}}\right)^{3{\beta}}}\right]\nonumber\\
&&\pm\frac{1}{8}\frac{\left[\left(\frac{{\beta}{c^2}}{2{\pi}G({t_s}-t)^3}-\frac{3{\rho_{m0}}{\beta{c^2}}}{{{a_0}^3}(t_s-t)}{\left(\frac{{t_s}-{t_0}}{{t_s}-{t}}\right)^{3{\beta}}}\right){{M_{S.I.}}^4}\right]}{\left[(\frac{{\beta}{c^2}}{4{\pi}G({t_s}-t)^2}-\frac{{\rho_{m0}}{{c^2}}}{{{a_0}^3}}{\left(\frac{{t_s}-{t_0}}{{t_s}-{t}}\right)^{3{\beta}}}+\frac{{M_{S.I.}}^4}{4}){{M_{S.I.}}^4}\right]^{\frac{1}{2}}}\label{rescale1}
\end{eqnarray}
which is nothing but the R.H.S. of (\ref{eqm1}).
A comparision of (\ref{rescale}) with (\ref{rescale1}) shows that
(\ref{eqm1}) is satisfied. Hence the reconstructed potential is consistent with the equation of motion of the scalar field.
%%%%%%%%%%%%%%%%%%%%%%% new inclusion finished

%%%%%%%%%%%%%%%%%%%%%%%%
\subsection{Input from the Observational data}
%%%%%%%%%%%%%%%%%%%%%%%%
We take into account the combined cosmic microwave background (CMB), baryon acoustic oscillations (BAO) and observational Hubble data ($H_0$) as well as the CMB-WMAP7 dataset seperately. The relevant results are tabulated in TABLE \ref{datatable}. The usual density parameter is $\Omega_{m}=8{\pi}G\rho_{m}/(3{H^2})$ and it is assumed to contain the baryonic matter $\Omega_{b}$ and cold dark matter $\Omega_{CDM}$ parts:
$\Omega_{m}= \Omega_{b}+\Omega_{CDM}$.
Using the expression of the critical density $\rho_{c} =\frac{3H^2}{8{\pi}G}$, the matter density at the present time can be found as $\rho_{m0}= {\Omega_{m0}}{\rho_{c0}}.$ We also set $a_0$ to 1.

From the phantom power law we get
\begin{eqnarray}
\beta=-{H_0}(t_s-t_0)
\end{eqnarray}
Assuming a flat geometry and that at late times the phantom dark energy will dominate the universe $t_s$ can be expressed as \cite{phantom2}
\begin{eqnarray}
{t_s}\simeq{t_0} +\frac{2}{3}|1 + w_{DE}|^{-1}{{H_0}^{-1}}(1-\Omega_{m0})^{-\frac{1}{2}} 
\end{eqnarray}
In deriving the above formula it has been assumed that at late times the
dark energy EoS parameter $w_{DE}$ approaches a constant value. The values of the derived parameters are given in TABLE \ref{resulttable}.
\newpage
%%%%%%%%%%%TABLE1%%%%%%%%%%%%%%%%%%%%%%%%%%%%%%%%%%%%%%%%%%%
\begin{table*}[!]
\begin{ruledtabular}
\begin{tabular}{ccc}
\textbf{Parameter} & \textbf{WMAP7+BAO+$H_0$} & \textbf{WMAP7}\\
\hline\hline
$t_0$ & $13.78\pm0.11$ Gyr [$(4.33\pm0.04) \times 10^{17}$ sec] & $13.71\pm0.13$  Gyr  [$(4.32\pm0.04) \times 10^{17}$ sec] \\
$H_0$ & $70.2^{+1.3}_{-1.4}$ km/s/Mpc & $71.4\pm2.5$ km/s/Mpc\\
$\Omega_{b0}$ & $0.0455\pm0.0016$ & $0.0445\pm0.0028$\\
$\Omega_{\mathrm{CDM}0}$ & $0.227\pm0.014$ & $0.217\pm0.026$\\
\end{tabular}
\caption{Maximum likelihood values  for the observed cosmological parameters in 1$\sigma$
confidence level
%the present time $t_0$, the present Hubble parameter $H_0$, the present baryon density parameter $\Omega_{b0}$ and the present cold dark matter density parameter $\Omega_{\mathrm{CDM}0}$, for WMAP7 as well as for the combined fitting WMAP7+BAO+$H_0$. The values are taken from
\cite{Komatsu}.} \label{datatable}
\end{ruledtabular}
\end{table*}
%%%%%%%%%%%%%%%%%%%%%%%%%%%TABLE2%%%%%%%%%%%%%%%%%%%%%%%%%%%%%%
\begin{table*}[!tbp]
\begin{ruledtabular}
\begin{tabular}{ccc}
    \textbf{Parameter} & \textbf{WMAP7+BAO+$H_0$} & \textbf{WMAP7}\\
    \hline \hline
$\beta$ & $-6.51^{+0.24}_{-0.25}$ & $−6.5\pm0.4$\\
\vspace{1mm}
$\rho_{m0}$ & $2.52^{+0.25}_{-0.24}\times{10^{-27}}kg/{m^3}$ & $2.50^{+0.48}_{-0.42}\times{10^{-27}}kg/{m^3}$\\
\vspace{1mm}
$\rho_{c0}$ & $9.3^{+0.3}_{-0.4}\times{10^{-27}}kg/{m^3}$ & $9.58^{+0.68}_{-0.66}\times{10^{-27}}kg/{m^3}$
\\

$t_s$ & $104.5^{+1.9}_{-2.0}$Gyr$[(3.30\pm0.06)\times{10^{18}}$ sec]
 & $102.3\pm3.5$Gyr$[(3.23\pm0.11)\times{10^{18}}$ sec]\\
\end{tabular}
\caption{Corresponding maximum likelihood values of the derived parameters
%power-law exponent $\beta$, the present matter energy density value $\rho_{m0}$, the present critical energy density value $\rho_{c0}$ and the Big Rip time $t_s$, for WMAP7 as well as for the combined fitting WMAP7+BAO+$H_0$
.}
\label{resulttable}
\end{ruledtabular}
\end{table*}

%We begin by fixing the value of the parameter $M$ by using the reality of the potential energy function (\ref{V_f}) where data from TABLE(\ref{datatable}) and TABLE(\ref{resulttable}) has been used. It is found that $M = 1 {\rm ev}$ is a good choice.
We are now almost in a position to calculate numerical values of various quantities as function of time. But one last point is still missing. We require to fix the parameter $M$ in the ghost-condensate model. The value of this parameter should be chosen so that the quantity within square root in (\ref{V_f}) and (\ref{KE_f}) is positive ensuring real values for the potential and kinetic energies. We find that $M = 1 {\rm ev}$ is a good choice. Also we choose the upper sign in (\ref{V_f}) in order to ensure positive potential energy. As a consequence the upper sign in equation (\ref{KE_f}) is selected (see the discussion under equation(\ref{KE1})).

Using values from TABLE \ref{datatable} and TABLE \ref{resulttable} in equations (\ref{V_f}) we get expressions of $V(t)$ for the WMAP7+BAO+$H_0$ dataset as:
\begin{eqnarray}
V(t)= && \left[\frac{6.823\times{10^{-9}}}{(3.3-t)^2}+\frac{5.24\times{10^{-10}}}{(3.3-t)^2}-\frac{0.661\times{10^{-19}}}{(3.3-t)^{-19.53}}-8.61\times10^{-9}\right] \nonumber\\
&& +\frac{1}{2}\left[\left(-\frac{6.987\times{10^{-10}}}{(3.3-t)^2}-\frac{2.642\times{10^{-19}}}{(3.3-t)^{-19.53}}+1.72\times10^{-8}\right)1.72\times10^{-8}\right]^\frac{1}{2}
\end{eqnarray}
and for the  WMAP7 dataset as:
\begin{eqnarray}
V(t)= && \left[\frac{6.802\times{10^{-9}}}{(3.23-t)^2}+\frac{5.232\times{10^{-10}}}{(3.23-t)^2}-\frac{1.088\times{10^{-19}}}{(3.23-t)^{-19.5}}-{8.61\times10^{-9}}\right] \nonumber\\
&& + \frac{1}{2}\left[(-\frac{6.976\times{10^{-10}}}{(3.23-t)^2}-\frac{4.352\times{10^{-19}}}{(3.23-t)^{-19.5}}+1.72\times10^{-8})1.72\times10^{-8}\right]^\frac{1}{2}
\end{eqnarray}
Similarly for the kinetic energy term we get from equation (\ref{KE_f}) 
\begin{eqnarray}
{\dot\phi^2} = 3.444\times10^{-8}-\left[(-\frac{6.987\times{10^{-10}}}{(3.3-t)^2}-\frac{2.642\times{10^{-19}}}{(3.3-t)^{-19.53}} + 1.72\times10^{-8})6.89\times10^{-8}\right]^\frac{1}{2}
\label{KE_ult1}
\end{eqnarray}
and 
\begin{eqnarray}
{\dot\phi^2} = 3.444\times10^{-8} -\left[(-\frac{6.976\times{10^{-10}}}{(3.23-t)^2}-\frac{4.352\times{10^{-19}}}{(3.23-t)^{-19.5}} + 1.72\times10^{-8})6.89\times10^{-8}\right]^\frac{1}{2} 
\label{KE_ult2}
\end{eqnarray}
for the WMAP7+BAO+$H_0$ and WMAP7 dataset respectively.

\begin{figure}[ht]
\begin{center}
\includegraphics[width=12cm, height=10cm, angle=270]{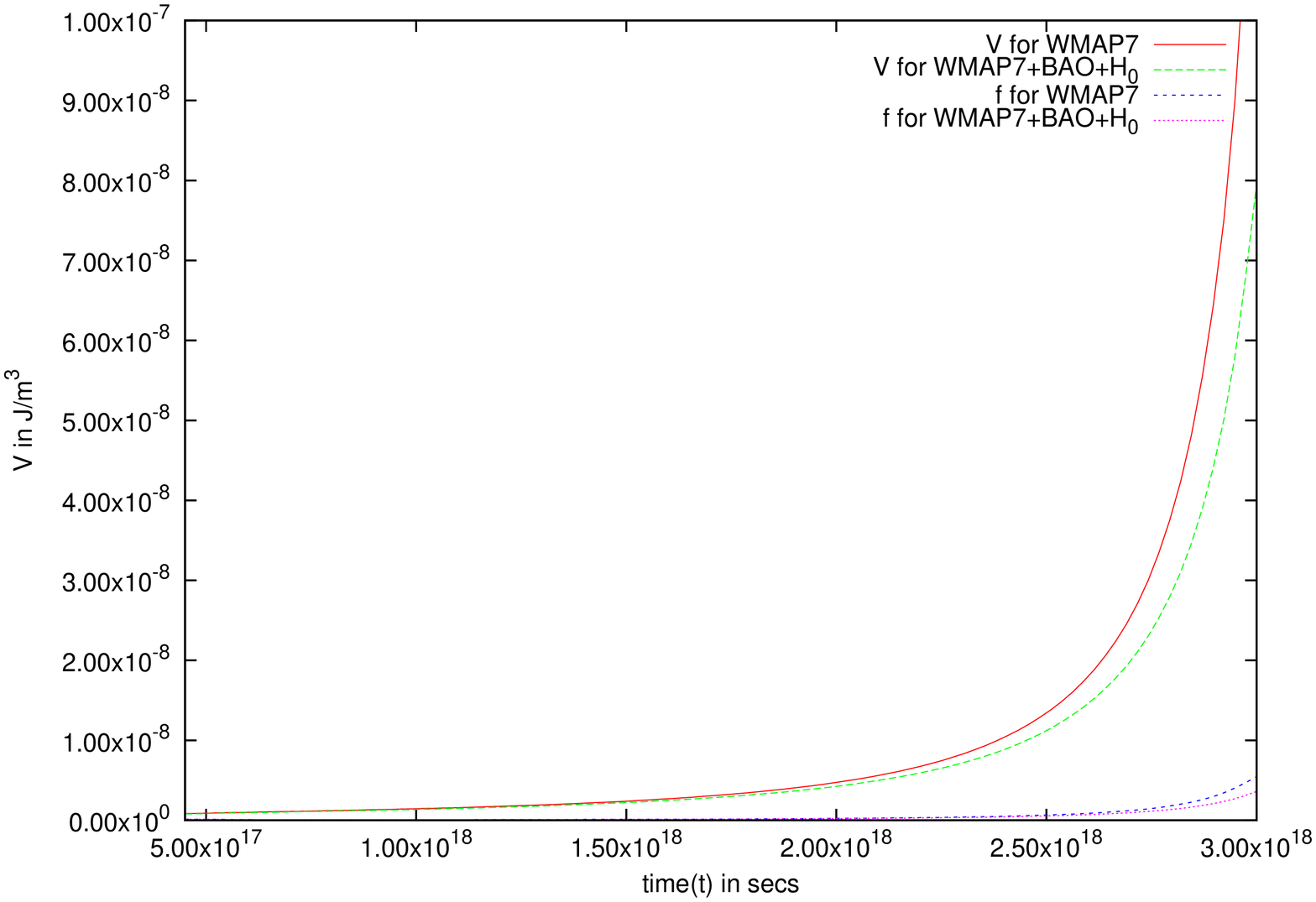}
\end{center}
\caption{{{The potential $V(\phi)$  plotted against time for the WMAP7 and WMAP7+BAO+$H_0$ data sets. The plot of the function $f({\dot{\phi}})$ vs time is also shown.}} }\label{V_plots}
%\caption{{\it{The potential $V(\phi)$ and the function $f({\dot{\phi}}^2)$ obtained from observational data fitting of WMAP7 and WMAP7+BAO+$H_0$ and plotted against time on the same graph.}} }\label{V_plots}
\end{figure}

\begin{figure}[ht]
\begin{center}
\includegraphics[width=12cm, height=10cm, angle=270]{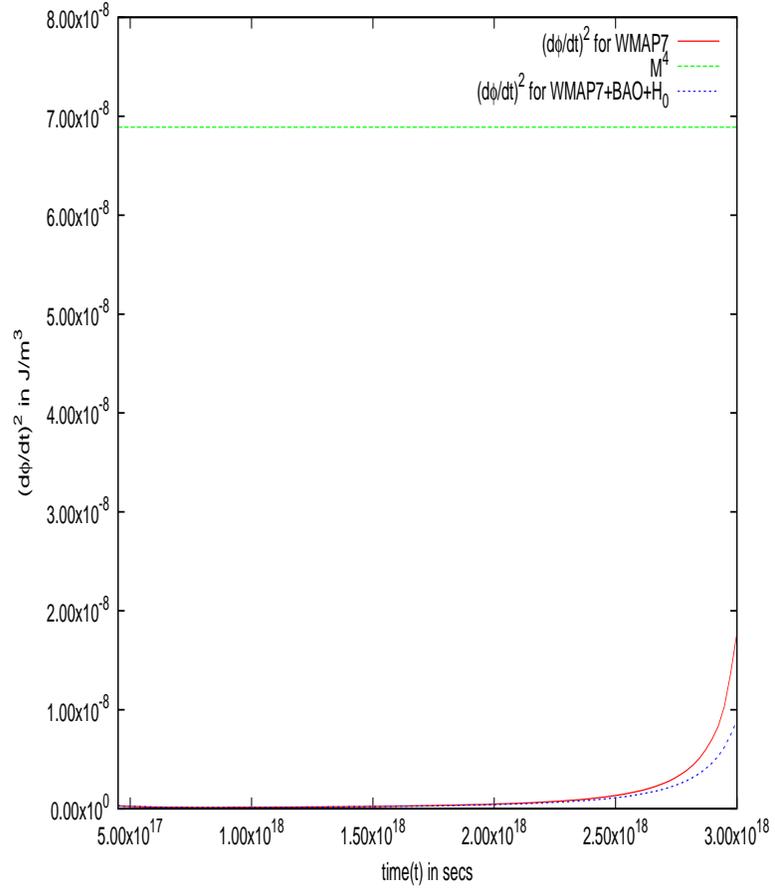}
\end{center}
\caption{{{${\left(\frac{d\phi}{dt}\right)}^2$ plotted against t for the WMAP7 and WMAP7+BAO+$H_0$ data set. Note that  $M^4 = 6.89e-08 J/m^3$ always lies above $\dot{\phi}^2$.}} }\label{KE_plots}
\end{figure}

In Figs \ref{V_plots} and  \ref{KE_plots} the evolution of the potential energy and the kinetic energy term are shown graphically against time. As expected, these quantities shoot up as the big rip time is approached. Superposed on the plot of potential energy in Fig. \ref{V_plots} is the funtion $f(\dot{\phi})$. It can be clearly seen that the condition (\ref{cond_1}) is satisfied throughout the future evolution. Again from Fig. \ref{KE_plots} we observe that the condition (\ref{cond_2}) is also satisfied because $\dot{\phi}^2$ always lies below $M^4= 6.89 \times 10^{-08}$. 
Thus we see that the criteria of realising the phantom regime (\ref{cond_1}, \ref{cond_2}) are satisfied by the present model.

%%%%%%%%%%%%%%%%%%%%%%%%%%%%%%%%%%%%%%%%%%%%%%%%%%%%%%%%%%%%%%%%%%%
\section{Conclusion}
%%%%%%%%%%%%%%%%%%%%%%%%%%%%%%%%%%%%%%%%%%%%%%%%%%%%%%%%%%%%%%%%%%%
Recent observations \cite{Komatsu, PANSTARR} indicate that there is a fair possibility of the late-time universe to follow the phantom evolution. The ghost condensate (GC) model is a dark energy model which realises the samke phantom evolution while eradicating some of the critical problems of the original phantom model. The inclution of a self-interaction in this model appears to be a matter of choice in the literature \cite{DE}. In this paper we have considered a ghost condensate (GC) model with an arbitrary potential term in a flat FLRW universe. The standard barotropic matter equation of state is assumed. Keeping the potential arbitrary we have derived new conditions for this model to realise the phantom regime. These include a condition on the potential energy (coming from the positive energy condition) and another condition on the allowed range of the kinetic energy so that the EoS parameter satisfies the phantom limit. This computation shows that the inclusion of a generic self-interaction widens the range of kinetic energy for achieving the phantom evolution. Naturally the question comes whether these new conditions derived here are maintained throughout the late time evolution of the universe. Now one has to start with a definite potential to trace the dynamics of any system. Since the purpose of the present paper is to stress the inclusion of a potential in the GC model we do not assume any specific functional form of the potential apriori.
We observed that the structure of the ghost-condensate model gives a non-trivial significance to the obvious identity (\ref{identity}). This allowed us to express the arbitrary potential in terms of the observable geometric quantities. These geometric quantities are model independent \cite{sahni} and are determined by observations.

\begin{figure}[ht]
\begin{center}
\includegraphics[width=12cm, height=10cm, angle=270]{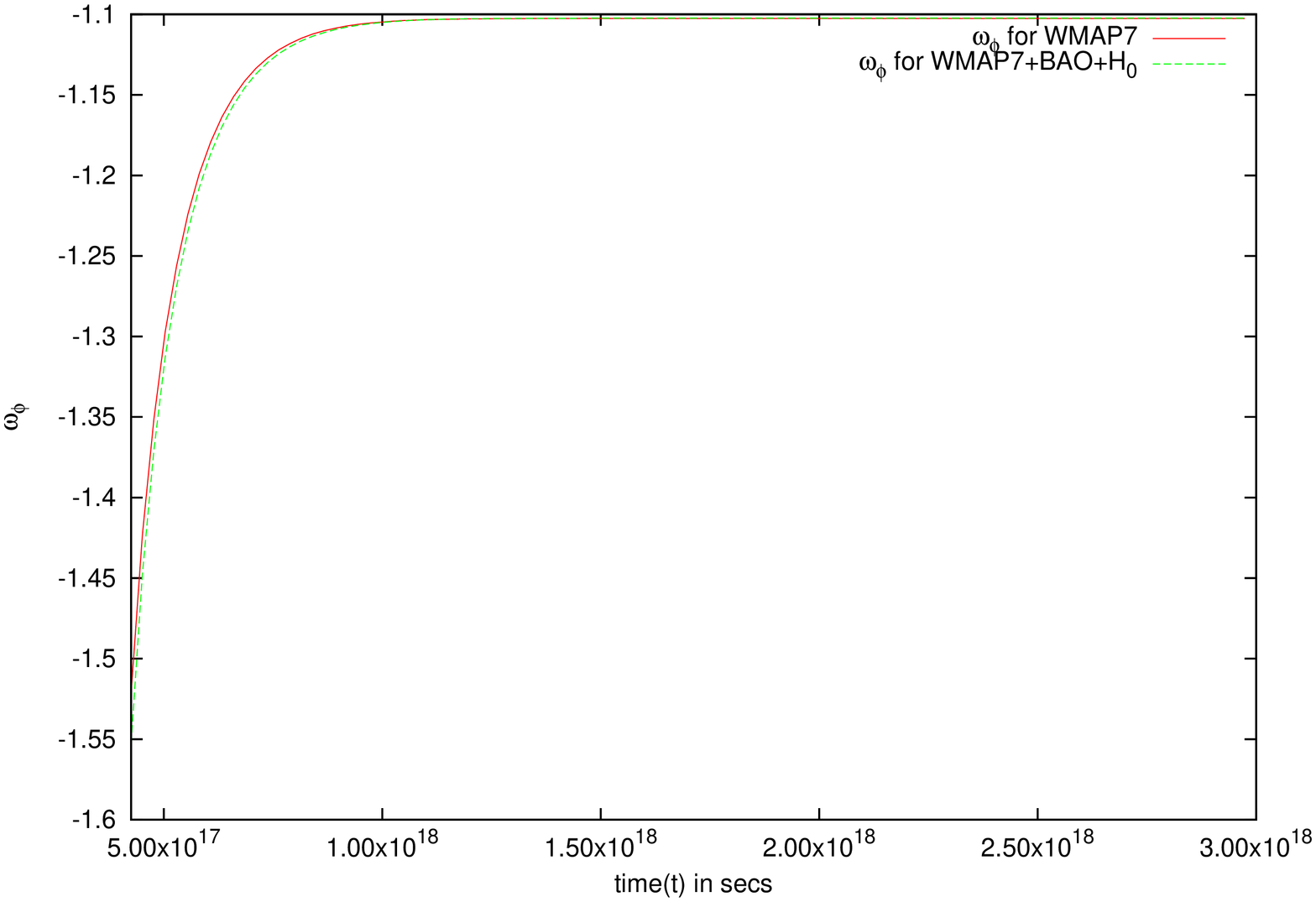}
\end{center}
\caption{{{${\omega_{\phi}}$ plotted against t for the WMAP7 and WMAP7+BAO+$H_0$ data set.}}}\label{Omegaplots}
\end{figure}

As recent observations \cite{Komatsu, PANSTARR} indicate that there is a fair possibility of the late-time universe to follow the phantom evolution we have assumed the phantom power law for the scale factor. To determine the parameters appearing in the power law we have used the combined WMAP7+BAO+Hubble as well as WMAP7 dataset.
Consequently, we obtained the potential and kinetic energies as functions of time. We have plotted the function $f\l(\dot{\phi}\r)$ (see equation (\ref{cond_1})) along with $V\l(\phi\r)$ in Fig. \ref{V_plots} and the line $\dot{\phi}{}^{2} = M^{4}$ along with $\dot{\phi}{}^{2}$ in Fig. \ref{KE_plots}. These plots revealed that the conditions which were derived earlier for our model to realise the phantom regime holds throughout the future evolution. Naturally the equation of state parameter $\omega_{\phi}$ should remain phantom-like. The variation of $\omega_{\phi}$ against time, shown in Fig. \ref{Omegaplots}, clearly exibits the same.
%%%%%%%%%%%%%%%%%%
\section*{Acknowledgement}
The authors thank the referee for his useful comments.
AS acknowledges the support by DST SERB under Grant No. SR/FTP/PS-208/2012.

%%%%%%%%%%%%%%%%FIGURES%%%%%%%%%%%%%%%%%%
\newpage
%%%%%%%%%%%%%%%%%%%%%%%%%%%%%%%%%%%%%%%%%%%%

\begin{thebibliography}{99}
\bibitem{NL1}A.~G.~Riess {\it et al.}  [Supernova Search Team Collaboration], Astron.\ J.\  {\bf 116}, 1009 (1998).

\bibitem{NL2}S. Perlmutter {\it et al.} [Supernova Cosmology Project Collaboration], Astrophys. J. {\bf 517}, 565 (1999).

\bibitem{wein}S.~Weinberg, Rev. Mod. Phys. {\bf 61} 1 (1989).

\bibitem{DE}Luca Amendola, Shinji Tsujikawa, {\it Dark Energy: Theory and Observations}, Cambridge University Press, 2010.

\bibitem{bamba} K.~Bamba, S.~Capozziello, S.~Nojiri, S.~D.~Odintsov, Astrophysics and Space Science (2012) 342:155-228.

\bibitem{quint1}B. Ratra and P.J.E. Peebles, Phys. Rev. {\bf D 37} 3406 (1988).

\bibitem{quint2} R.R. Caldwell, R. Dave, and P.J. Steinhardt, Phys. Rev. Lett. {\bf 80} 1582 (1998).

\bibitem{quint3} S.M. Carroll, Phys. Rev. Lett. {\bf 81} 3067 ((1998)).

\bibitem{quint4} I. Zlatev, L.M. Wang, and P.J. Steinhardt, Phys. Rev. Lett. {\bf 82} 896 (1999).

\bibitem{quint5} P.J. Steinhardt, L.M. Wang, and I.M. Zlatev, Phys. Rev. {\bf D 59} 123504 (1999).

\bibitem{quint6} A. Hebecker, and C. Wetterich, Phys. Lett. {\bf B 497} 281 (2001).

\bibitem{quint7} R.R. Caldwell, and E.V. Lindner, Phys. Rev. Lett. {\bf 95} 141301 (2005).

\bibitem{kessence1} T. Chiba, T. Okabe, and M. Yamaguchi, Phys. Rev. {\bf D 62} 023511 (2000).

\bibitem{kessence2} C. Armandariz-Picon, V.F. Mukhanov, and P.J. Steinhardt, Phys. Rev. Lett. {\bf 85} 4438 (2000); ibid, Phys. Rev. {\bf D 63} 103510 (2001).

\bibitem{kessence3}Y. Kamenshchik, U. Moschella, and V. Pasquier, Phys. Lett. {\bf B 511} 265 (2001).

\bibitem{Komatsu} E. Komatsu {\it et al}, Astrophys.J.Suppl.192:18,2011 

\bibitem{PANSTARR}A.~Rest {\it et.al.}, Astrophys.J. 795 (2014) 1, 44, arXiv:1310.3828 [astro-ph.CO]

\bibitem{phantom1} R.~R.~Caldwell, M.~Kamionkowski and N.~N.~Weinberg, Phys.\ Rev.\ Lett.\  {\bf 91} 071301 (2003).

\bibitem{phantom2} R. R. Caldwell, Phys. Lett. {\bf B 545} 23 (2002).

\bibitem{phantom3}  S. Nojiri and S. D. Odintsov, Phys. Lett. {\bf B 562} 147 (2003).

\bibitem{phantom4} P.~Singh, M.~Sami and N.~Dadhich, Phys.\ Rev.\  D {\bf 68} 023522 (2003).

\bibitem{phantom5} J.~M.~Cline, S.~Jeon and G.~D.~Moore, Phys.\ Rev. {\bf D 70} 043543 (2004).

\bibitem{phantom6} V. K. Onemli and R. P. Woodard, Phys.\ Rev.{\bf D 70} 107301 (2004).

\bibitem{phantom_obs1} M.~Sethi, A.~Batra and D.~Lohiya, Phys. Rev. {\bf D 60} 108301 (1999).

\bibitem{phantom_obs2} M.~Kaplinghat, G.~Steigman and T.~P.~Walker, Phys. Rev. {\bf D 61} 103507 (2000).

\bibitem{phantom_obs3} M.~Kaplinghat, G.~Steigman, I.~Tkachev and T.~P.~Walker, Phys. Rev. {\bf D 59}, 043514 (1999).

\bibitem{phantom_obs4} D.~Lohiya and M.~Sethi, Class. Quan. Grav. {\bf 16}, 1545 (1999).

\bibitem{phantom_obs5} G.~Sethi, A.~Dev and D.~Jain, Phys.\ Lett. {\bf B 624}, 135 (2005).

\bibitem{phantom_obs6} S.~W.~Allen, R.~W.~Schmidt, A.~C.~Fabian, Mon. Not. Roy. Astro. Soc. {\bf 334}, L11 (2002).

\bibitem{GC} N.~Arkani-Hamed, H.~C.~Cheng, M.~A.~Luty, and S.~Mukohyama, JHEP {\bf 0405} (2004), 074.

\bibitem{gumjudpai} C.~Kaeonikhom, B.~Gumjudpai, E.~N.~Saridakis, Phys. Lett. {\bf B 695} 45, 2011.

\bibitem{sahni}V.~Sahni, T.~Saini, A.~A.~Starobinsky, U.~Alam, JETP Lett.77:201-206,2003; Pisma Zh.Eksp.Teor.Fiz.77:249-253,2003



%\bibitem{ekpyrotic1}J.~Khoury, B.~A.~Ovrut, P.~J.~Steinhardt, and N.~Turok, Phys. Rev. {\bf D 64} 123522 (2001).
%, [arXiv.org: hep-th/0103239].

%\bibitem{ekpyrotic2} J.~Khoury, B.~A.~Ovrut, P.~J.~Steinhardt, and N.~Turok, Phys. Rev. {\bf D 66} 46005 (2002).
%, [arXiv.org: hep-th/0109050].

%\bibitem{ekpyrotic3}J.~L.~Lehners, Phys. Rept. {\bf 465}, 223 (2008).
%, [arXiv.org: 0806.1245].

%\bibitem{bounce1}J.~Khoury, B.~A.~Ovrut, N.~Seiberg, P.~J.~Steinhardt, and N.~Turok, Phys. Rev. {\bf D 65} 86007 (2002).
%, [arXiv.org: hep-th/0108187].

%\bibitem{bounce2}P.~Creminelli and L.~Senatore, JCAP 0711, 010 (2007).
%, [hep-th/0702165].

%\bibitem{bounce3}C.~Lin, R.~H.~Brandenberger, and L.~Levasseur~Perreault, JCAP {\bf 1104} 019 (2011).
%, [arXiv.org:1007.2654].

%\bibitem{bounce4}Y.-F.~Cai, D.~A.~Easson, and R.~Brandenberger, JCAP {\bf 1208}, 020 (2012).
%\bibitem{buchbinder} E. I. Buchbinder, J. Khoury, and B. A. Ovrut, Phys. Rev. D 76, 123503 (2007), [arXiv.org: hep-th/0702154; E. I. Buchbinder, J. Khoury, and B. A. Ovrut, JHEP 11, 076 (2007), [arXiv.org: 0706.3903 [hep-th]]; E. I. Buchbinder, J. Khoury, and B. A. Ovrut, Phys. Rev. Lett. 100, 171302 (2008), [arXiv.org: 0710.5172 [hep-th]].
\end{thebibliography}
\end{document}